%% file: 00_main.tex
\documentclass[conference]{IEEEtran}
\IEEEoverridecommandlockouts
\usepackage{cite}
\usepackage{amsmath,amssymb,amsfonts}
\usepackage{algorithmic}
\usepackage{graphicx}
\usepackage{textcomp}
\usepackage{xcolor}
\usepackage[a4paper, total={184mm,239mm}]{geometry}
\def\BibTeX{{\rm B\kern-.05em{\sc i\kern-.025em b}\kern-.08em
    T\kern-.1667em\lower.7ex\hbox{E}\kern-.125emX}}
    
\usepackage{algorithm}
\usepackage{subcaption}
\usepackage{booktabs}
\usepackage{mathtools}
\usepackage{balance}
\usepackage{setspace}
\usepackage{multirow}
\usepackage{caption}
\usepackage{listings}
\usepackage{comment}
\usepackage{fancyhdr}

\lstdefinelanguage{CUDA}%
  {morekeywords={__global__, __device__, __shared__, __host__, __constant__, int, const, if, else, for, return},%
   morecomment=[l]{//},%
   morecomment=[s]{/*}{*/},%
   morestring=[b]",%
   sensitive=true%
  }[keywords, comments, strings]

\usepackage{xpatch}
\xpatchcmd{\algorithmic}{\setcounter}{\algorithmicfont\setcounter}{}{}
\providecommand{\algorithmicfont}{}
\providecommand{\setalgorithmicfont}[1]{\renewcommand{\algorithmicfont}{#1}}

\setalgorithmicfont{\footnotesize}
\begin{document}

\title{\emph{SparseInfer}: Training-free Prediction of Activation Sparsity for Fast LLM Inference\\
}

\author{\IEEEauthorblockN{Jiho Shin}
\IEEEauthorblockA{
\textit{University of Seoul}\\
Seoul, South Korea \\
sjh010529@uos.ac.kr}
\and
\IEEEauthorblockN{Hoeseok Yang$^{*}$\thanks{$^{*}$Youngmin Yi and Hoeseok Yang are corresponding authors.}}
\IEEEauthorblockA{
\textit{Santa Clara University}\\
Santa Clara, CA, USA \\
hoeseok.yang@scu.edu}
\and
\IEEEauthorblockN{Youngmin Yi$^{*}$\thanks{This work was supported by Institute of Information communications Technology Planning Evaluation (IITP) grant funded
by the Korea govern- ment(MSIT) (No.2022-0-00498)}}
\IEEEauthorblockA{
\textit{Sogang University}\\
Seoul, South Korea \\
ymyi@sogang.ac.kr}
}

\maketitle
\thispagestyle{fancy}

\input{0_abstract.tex}


\input{1_intro.tex}

\input{2_related_work.tex}

\input{3_background.tex}

\input{4_proposed.tex}
\input{6_experiments.tex}

\input{7_conclusion.tex}

\bibliography{references}
\bibliographystyle{IEEEtran}

\end{document}

%% file: 0_abstract.tex
\begin{abstract}

Leveraging sparsity is crucial for optimizing large language model (LLM) inference; however, modern LLMs employing SiLU as their activation function exhibit minimal activation sparsity. 
Recent research has proposed replacing SiLU with ReLU to induce significant activation sparsity and showed no downstream task accuracy degradation through fine-tuning.
However, taking full advantage of it required training a predictor to estimate this sparsity.
In this paper, we introduce \emph{SparseInfer}, a simple, light-weight, and training-free predictor for activation sparsity of ReLU-fied LLMs, in which activation sparsity is predicted by comparing only the sign bits of inputs and weights. 
To compensate for possible prediction inaccuracy, an adaptive tuning of the predictor’s conservativeness is enabled, which can also serve as a control knob for optimizing LLM inference. 
The proposed method achieves approximately 21\% faster inference speed over the state-of-the-art, with negligible accuracy loss of within 1\%p.
\end{abstract}

%% file: 1_intro.tex
\section{Introduction}
\label{sec:introduction}

Following the remarkable success of Generative Pre-trained Transformer (GPT) models~\cite{radford2018improving} based on the Transformer architecture~\cite{vaswani2017attention}, Large Language Models (LLMs) are being actively employed across a wide range of fields.
The inference of LLMs based on the GPT architecture first performs an initial forward pass, also known as \emph{prefill}, followed by an autoregressive \emph{decoding} process that generates tokens sequentially, one at a time. 
Due to this sequential nature, memory accesses occur frequently in the decoding stage. This can lead to I/O bottlenecks, which in turn limits the overall computational performance.
Well-known optimization methods such as pruning~\cite{ma2023llm} or quantization~\cite{yao2022zeroquant} have been applied to LLMs to reduce this memory access overhead but can result in an accuracy loss.

Another approach for optimizing LLM inference is to take advantage of activation sparsity. Unlike pruning techniques, which either remove zero-valued weights from the model or set some weights to zero, activation sparsity exploits the zeros that occur dynamically during runtime, depending on the input. A major source of this activation sparsity comes from activation layers. While Sigmoid Linear Unit (SiLU)~\cite{ramachandran2017searching} or Gaussian Error Linear Unit (GELU)~\cite{hendrycks2016gaussian} is popularly used as activation layers in the Multi-Layer Perceptron (MLP) of recent LLMs, they exhibit significantly lower sparsity compared to Rectified Linear Unit (ReLU). However, recent studies~\cite{mirzadeh2023relufication,song2024prosparse} have shown that replacing SiLU or GELU with the more sparse ReLU can maintain nearly the same level of accuracy with minor fine-tuning.

Although such \emph{ReLU-fied} LLMs have successfully increased the degree of activation sparsity, it is still not feasible to know in advance which inner-product operations will ultimately result in zeros after passing through the activation layer, thus can be skipped, at design-time. Therefore, the LLM inference needs to be assisted by a method that can effectively predict this during runtime. A recent study~\cite{liu2023dejavu} proposed a method that involves training an additional neural network, which consists of a couple of fully connected layers, per each MLP block to predict activation sparsity. However, this approach has the drawback of requiring a separate predictor to be trained. In addition, since the predictor needs to reside in memory throughout inference, the impact of this additional neural network on memory footprint, as well as its computational overhead, cannot be ignored.

In this paper, we propose \emph{SparseInfer}, which accelerates General Matrix-Vector Multiplication (GEMV) operations in the MLPs of ReLU-fied LLMs by leveraging activation sparsity. Unlike previous methods, the proposed approach efficiently predicts the sparsity pattern without requiring additional predictor training, making it easily applicable to legacy pre-trained LLMs. Furthermore, this technique is portable across various types of hardware, including both commercial off-the-shelf (COTS) processors such as GPUs and CPUs, as well as customized hardware accelerators, as it only requires a simple lookup of partial information about the elements involved in the inner product of GEMV operations. 
We validate our implementation using a mobile-grade GPU. 

The proposed activation sparsity prediction method is based on approximating the sign of the input to ReLU without calculating the exact result. Specifically, the sign of each element in the two vectors involved in the inner product is compared with the corresponding element's sign in the other vector using XOR, predicting the sign of each element-wise multiplication result. Then, assuming the absolute values of the elements in the two vectors follow a normal distribution, we compare the number of negative and positive results before accumulation to predict whether the input to ReLU will be positive or negative. If more element-wise products are predicted to be negative, the final result is likely to be negative. Since a negative value would result in zero after passing through ReLU, this is assumed to contribute to activation sparsity.

Despite a high level of accuracy, this predictor is inevitably subject to a certain degree of inaccuracy. Experimental results indicate that this inaccuracy is relatively more noticeable in the early layers. To overcome this, two additional measures have been implemented. The first measure is to consider additional sparsity identified during MLP execution, which the predictor initially missed, as \emph{actual} sparsity (as opposed to \emph{predicted} sparsity). Secondly, we extract a tunable parameter to adaptively control the level of conservativeness in predictions. With this, we can explore the trade-off between inference speed and potential accuracy degradation. Also, this can serve as an important control knob for design space exploration (DSE) in optimizing LLM inference,
given the target platform, the model, and the downstream task.

\emph{SparseInfer} was implemented on an open source LLM inference engine, \emph{llama.cpp}~\cite{llama_cpp}. The proposed XOR-based activation sparsity prediction method was implemented as a CUDA kernel to enable parallel processing on the GPU. Based on this prediction, the MLP operations were optimized to exploit activation sparsity. To further enhance the performance, kernel fusion was applied to the MLP and activation layers, reducing memory access overhead. Experiments on NVIDIA Jetson Orin AGX 64GB show that the overhead of the predictor was reduced by a factor of four compared to the existing activation sparsity prediction method, and the inference speed per token with the Llama2 13B and 7B models improved by approximately 21\% and 19\% compared to the state-of-the-art, while maintaining the accuracy loss within 1\%p.

%% file: 2_related_work.tex
\section{Related Work}
\label{sec:related_work}
\textbf{Activation Sparsity-induced LLM}:
Mirzadeh et al.\cite{mirzadeh2023relufication} proposed a technique called ReLUfication which replaces activation functions in LLM, such as SiLU, with ReLU. They showed ReLUfication can induce significant activation sparsity while accuracy loss is negligible after fine-tuning. 
\emph{ProSparse}~\cite{song2024prosparse} significantly induced activation sparsity by progressively applying sparsity regularization after Relufication. It also sets a positive value instead of zero for a ReLU threshold and achieves larger sparsity using a technique called \emph{FATReLU}~\cite{kurtz2020inducing}. 
It induced a 22\%p higher sparsity compared to \emph{ReluLLaMa}~\cite{relullama}, an open-source ReLU-based LLM fine-tuned from \emph{Llama2}, achieving 66\% sparsity. Despite a slight performance drop in some common sense benchmarks, the resultant model generally maintained comparable performance on average and even performed slightly better in certain cases, achieving up to a $4.65\times$ speedup in inference time with \emph{PowerInfer}~\cite{song2023powerinfer}. A similar approach, \emph{TurboSparse}~\cite{song2024turbo}, was proposed  for Mixtral models.

Another approach is to employ the SiLU activation function as is but set an input distribution-based threshold to induce sparsity\cite{song2024prosparse, liu2024teal}. While \emph{CATS}~\cite{song2024prosparse} only sparsified the Feed-Forward Network (FFN) layers, \emph{TEAL}\cite{liu2024teal} applied the method to the Attention layer as well. Compared to ReLUfication, this approach does not require fine-tuning but shows lower sparsity at the comparable performance; \emph{CATS} reported 15\% speedup in inference.

\textbf{Activation Sparsity Prediction}:
\emph{DEJAVU}~\cite{liu2023dejavu} proposed utilizing activation sparsity in the Attention and MLP blocks of a Transformer layer, leveraging a sparsity predictor composed of two fully connected layers. While this approach is argued to be less computationally intensive than a nearest-neighbor-based predictor, it still incurs considerable overhead. Furthermore, the predictor necessitates additional training.
Also, it is not easy to determine the training hyper-parameters, such as the dimension size of predictors in a layer, which may impact the prediction performance. It reduced the inference latency for \emph{OPT-175B} model by $2\times$ compared to \emph{FasterTransformer} on A100 GPUs.
\emph{PowerInfer}~\cite{song2023powerinfer} adopted \emph{DEJAVU}'s activation sparsity predictor and achieved a large speedup of up to $11.6\times$ over \emph{llama.cpp} when the GPU memory could not load the entire model. The breakdown of speedup showed that the speedup due to exploiting sparsity was about twofold.

\textbf{Novelty of the proposed method}:
Most existing methods for utilizing LLM activation sparsity, if not all, rely on predictors that require training. This approach has a fundamental drawback in that retraining is necessary whenever the model changes or its quantization is modified. The proposed method, in contrast, predicts sparsity dynamically and approximately without the need for training, distinguishing it from previous methods. The proposed predictor not only eliminates the need for additional training but also reduces memory and computational overhead. Although similar prediction methods have been explored in Convolutional Neural Networks (CNNs), such as in SeerNet~\cite{cao2019seernet}, it has not been applied to LLMs to the best of our knowledge.

%% file: 3_background.tex
\section{Background: Gate-Based MLP}
\label{sec:background}

In this section, we illustrate the operation of the MLP block, which we aim to optimize using activation sparsity and where the majority of time is spent during the LLM decoding process.\footnote{In our profiling of the inference of \emph{Llama2-13B} using \emph{llama.cpp} on NVIDIA Jetson Orin AGX 64GB, we observed that during decoding, self-attention and MLP computations accounted for 38\% and 62\%, respectively.} 
Due to the ease of presentation and space constraints, we focus our explanation on the Gate-based MLP used in the Llama model~\cite{touvron2023llama}. However, it is important to note that the proposed technique is not limited to this specific model and can be applied, without loss of generality, to any models that can be ReLU-fied. As demonstrated in Mirzadeh et al.~\cite{mirzadeh2023relufication}, ReLUfication can be effectively applied to different types of MLP blocks found in models such as \emph{Falcon}~\cite{almazrouei2023falcon} and \emph{OPT}~\cite{zhang2023opt}.

The gate-based MLP block consists of three linear layers ($W_{gate}$, $W_{up}$, and $W_{down}\in \mathbb{R}^{d\times k}$ in which $k>d$) with $d$ being the model dimension and one activation layer ($\sigma(\cdot)$). It can be formulated as follows:
\begin{equation}
\label{eq:mlp}
    MLP(X) =(\sigma(X\cdot W_{gate})\odot(X\cdot W_{up}))\cdot W_{down}^T,
\end{equation}
where $\odot$ denotes an element-wise multiplication between two vectors. Specifically, the input vector $X\in \mathbb{R}^{1\times d}$ goes through 4 steps. \begin{enumerate}
    \item (gate computation) The input $X$ is first used to compute the gate values. This is done by applying a linear transformation indicated by $W_{gate}$ followed by the activation function, thus formulated as $h1=\sigma(X\cdot W_{gate})$.
    \item (input processing) Simultaneously, the input vector $X$ undergoes a separate linear transformation by $W_{up}$, resulting in $h2=X\cdot W_{up}$. 
    \item (gate application) The gate values computed in step 1 are then applied element-wise to the processed input from step 2. This is achieved through element-wise multiplication ($h3=h1\odot h2$).
    \item (output generation) Finally, the gated vector is passed through another linear transformation using the weight matrix $W_{down}$, resulting in the final output as $h3\cdot W_{down}^T$.
\end{enumerate}

%% file: 4_proposed.tex
\section{Proposed Method: \emph{SparseInfer}}
\label{sec:proposed}

\begin{figure}[t]
\centering
\includegraphics[width=1\linewidth]
{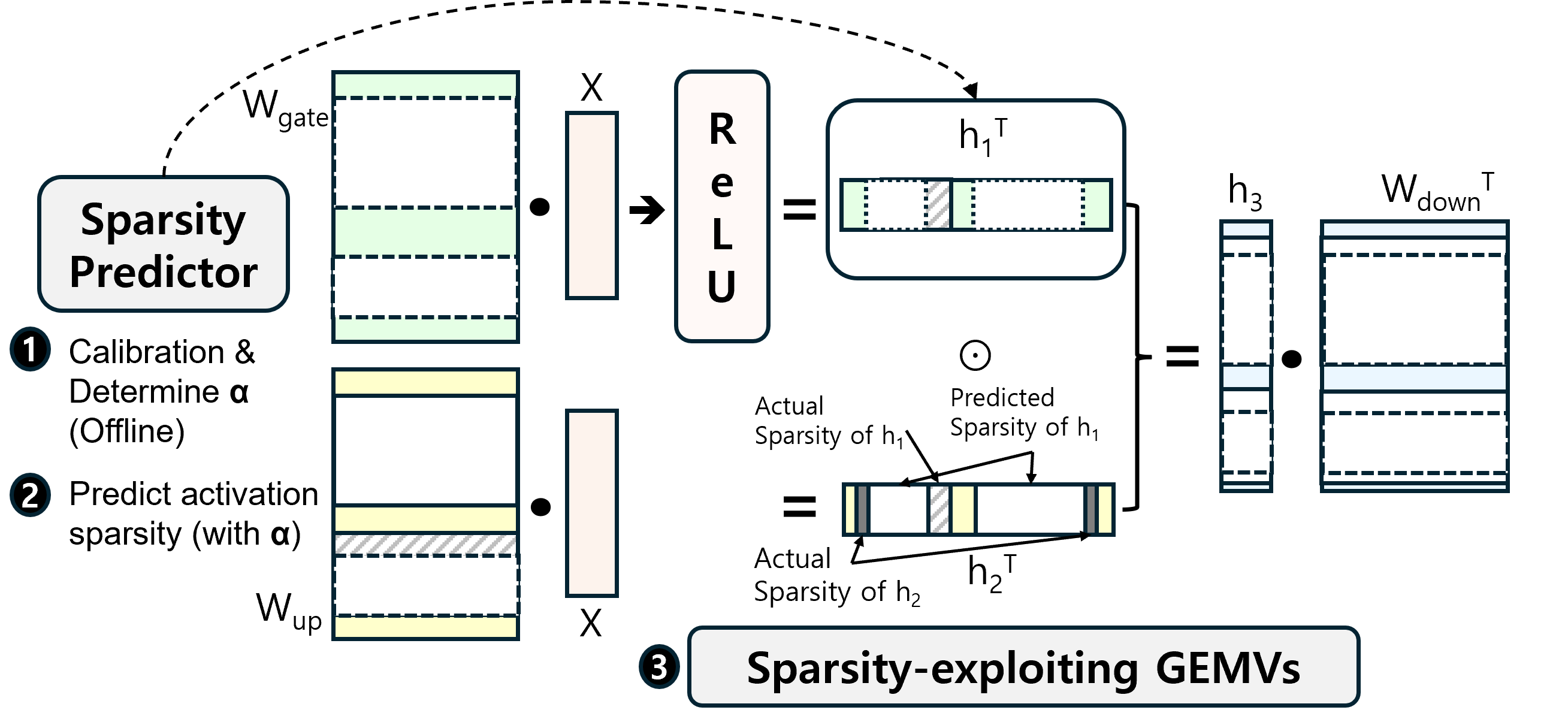}
\caption{Overview of SparseInfer approach}
\label{fig:overview}
\end{figure}

 As explained in Section~\ref{sec:background}, ReLU is applied to the result of Matrix-Vector multiplication of $W_{gate}$ and $X$, producing $h1$ vector. Note that the activation sparsity of each layer is approximately 90\%\cite{song2024prosparse}, and the corresponding rows of $W_{gate}$ matrix to the zero elements in $h1$ need not be loaded and computed. The same applies to $W_{up}$ matrix in step 2 as $h2$ is used for element-wise vector-vector multiplication with $h1$. Furthermore, the activation sparsity of the $h3$ vector which should be the same or higher than that of the $h1$ vector is exploited to skip the corresponding rows of the transposed $W_{down}$ matrix in step 3. These significantly reduce the inference time. 

Fig.~\ref{fig:overview} illustrates the overview of how the proposed approach, \emph{SparseInfer}, utilizes the activation sparsity. \textcircled{1} First, we set the coefficients for the predictor to efficiently predict the sparsity, which is done through an offline effort. \textcircled{2} Next, we predict the sparsity in $h1$ by only looking up partial information from $W_{gate}$ and $X$. \textcircled{3} Finally, based on the predicted results from step 2, we skip the memory loading and computations of steps 1-4 for the positions in $h1$ that are predicted to be zero.

Note that, step 2, which is found in \emph{Llama} but not in \emph{Falcon} and \emph{OPT} models, can be executed in parallel with step 1, e.g., using Concurrent Kernel Execution (CKE) in CUDA. 
Alternatively, executing steps 1 and 2 sequentially offers two advantages.
First, the steps can be implemented in a single kernel (i.e., kernel fusion), eliminating the store and the load of input vector $X$. 
A more significant advantage of executing steps 1 and 2 sequentially is the ability to compensate for the predictor's inaccuracy. The predictor relies solely on sign information, which can lead to inevitable inaccuracies. Therefore, to minimize accuracy loss, the prediction of activation sparsity, performed before step 1, is conducted conservatively, which will be explained in Section~\ref{subsec:prediction}.
After step 1, in addition to the predicted sparsity, the exact, actual sparsity that was identified can be utilized.
In other words, when steps 1 and 2 are executed sequentially, it becomes possible to account for the sparsity missed by the initial predictor in subsequent steps. 
This activation sparsity to which we coined the term \textit{actual sparsity}, in addition to the predicted sparsity, can be exploited in the subsequent steps in the MLP block. The actual sparsity of $h2$, as well as that of $h1$, is exploited in step 4.

\begin{figure*}[t]
\centering
\includegraphics[width=1\linewidth]
{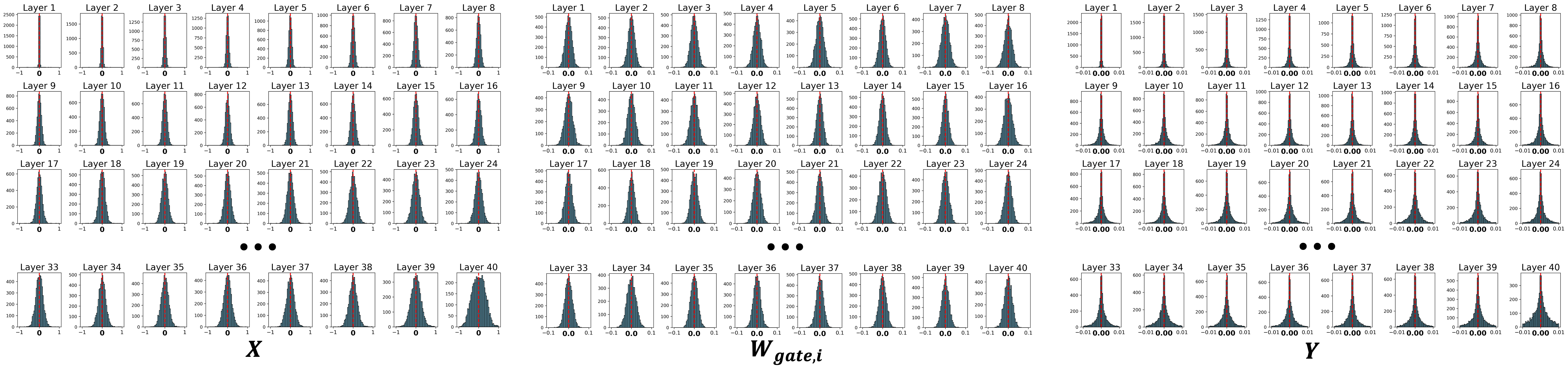}
\caption{Distribution of $X$, $W_{gate,i}$, and $Y=X\odot W_{gate,i}$ values obtained from \emph{ProSparse-Llama2-13B} during 8-shot inference on the \emph{GSM8K} dataset.}
\label{fig:dist}
\end{figure*}

\subsection{Training-free Sparsity Prediction}
\label{subsec:prediction}

\emph{SparseInfer} is designed to approximately predict activation sparsity without additional training. Since we target ReLU-fied LLMs, it is not necessary to predict the \emph{exact} value of the inputs to the ReLU. Instead, we only need to determine whether it is positive or negative. To this end, we devise a technique that minimizes memory access by using only the sign information of $W_{gate}$ and $X$. This allows us to approximate the sign information of the result of $X \cdot W_{gate}$ in step 1 (gate computation) will be positive or negative.

This approximation is relatively accurate because it relies on the assumption that the elements of $X$ and row vector of $W_{gate}$ ($W_{gate,i}$) follow independent Gaussian distributions with a mean equal to zero \cite{liu2024teal, tim2023qlora}. It is known that the products of two variables following independent Gaussian distributions with zero means have a symmetric distribution with a mean equal to zero\textcolor{black}{\cite{Craig1936OnTF}}. In this distribution, the ratio of positive to negative values is generally similar, meaning that the probability of individual products being positive or negative is almost equal. As a result, if there are more positive or negative values, the cumulative result is highly likely to have the sign of the more frequent values. This allows us to predict whether the result of $X \cdot W_{gate}$ will be positive or negative with relatively high accuracy, even without knowing the exact values.

Fig.~\ref{fig:dist} illustrates the distribution of values in a row vector of $W_{gate}$ ($W_{gate,i}$) and $X$ obtained from the \emph{ProSparse-Llama2-13B} model~\cite{song2024prosparse} when performing 8-shot inference on a subset of the \emph{GSM8K} dataset\cite{Karlgms8k}. Due to space constraints, intermediate layers exhibiting similar distributions have been omitted.

Both $X$ and $W_{gate}$ exhibit distributions that closely resemble Gaussian distributions, with an approximately equal ratio of positive and negative values. Moreover, their products $Y = X \odot W_{gate,i}$ follow a symmetric distribution with a mean approaching zero, as previously hypothesized. This confirms the validity of our initial assumption regarding the distributions of $X$ and $W_{gate}$, and consequently, the accuracy of our approximation method. In the early layers, the distribution of $X$ values is dominated by near-zero negative and near-zero positive values, leading to a narrow distribution heavily concentrated around zero, rather than following a typical normal distribution. This characteristic is believed to contribute to prediction inaccuracies.

The proposed prediction method estimates whether the result of the inner product between each row of $W_{gate}$, denoted as $W_{gate,i}$, and $X$ will become zero after passing through the ReLU function. By looking up only the most significant bits (MSBs) of $W_{gate,i}$ and $X$, we can predict the sign of each element-wise product through an XOR operation. Specifically, if the result of the XOR operation is 0, the product is predicted to be positive; if it is 1, the product is predicted to be negative. Based on this, we can determine the number of positive elements ($N_{pos}$) and negative elements ($N_{neg}$) in the resulting vector from the inner product of $W_{gate,i}$ and $X$. If $N_{pos} < N_{neg}$, it is likely that the accumulated result will be negative, and the corresponding element in $h1$ will be likely to be zero after passing through the ReLU function.

Since this prediction is not based on exact values but on empirical observations and statistical hypotheses, it comes with inevitable inaccuracies. To compensate for the potential inaccuracies that this may introduce into the inference process, a tunable coefficient $\alpha$ can be added to the equation, allowing for the adjustment of the aggressiveness or conservativeness of the activation sparsity prediction. Specifically, the above inequality is refined as follows:
\begin{equation}
\label{eq:th}
\alpha \cdot N_{pos} < N_{neg}   . 
\end{equation}
In this case, $\alpha>1.0$ results in a more conservative prediction of activation sparsity, whereas $\alpha<1.0$ leads to a more aggressive prediction, increasing the degree of activation sparsity.

\begin{figure}[tb]
\centering
\includegraphics[width=\linewidth]
{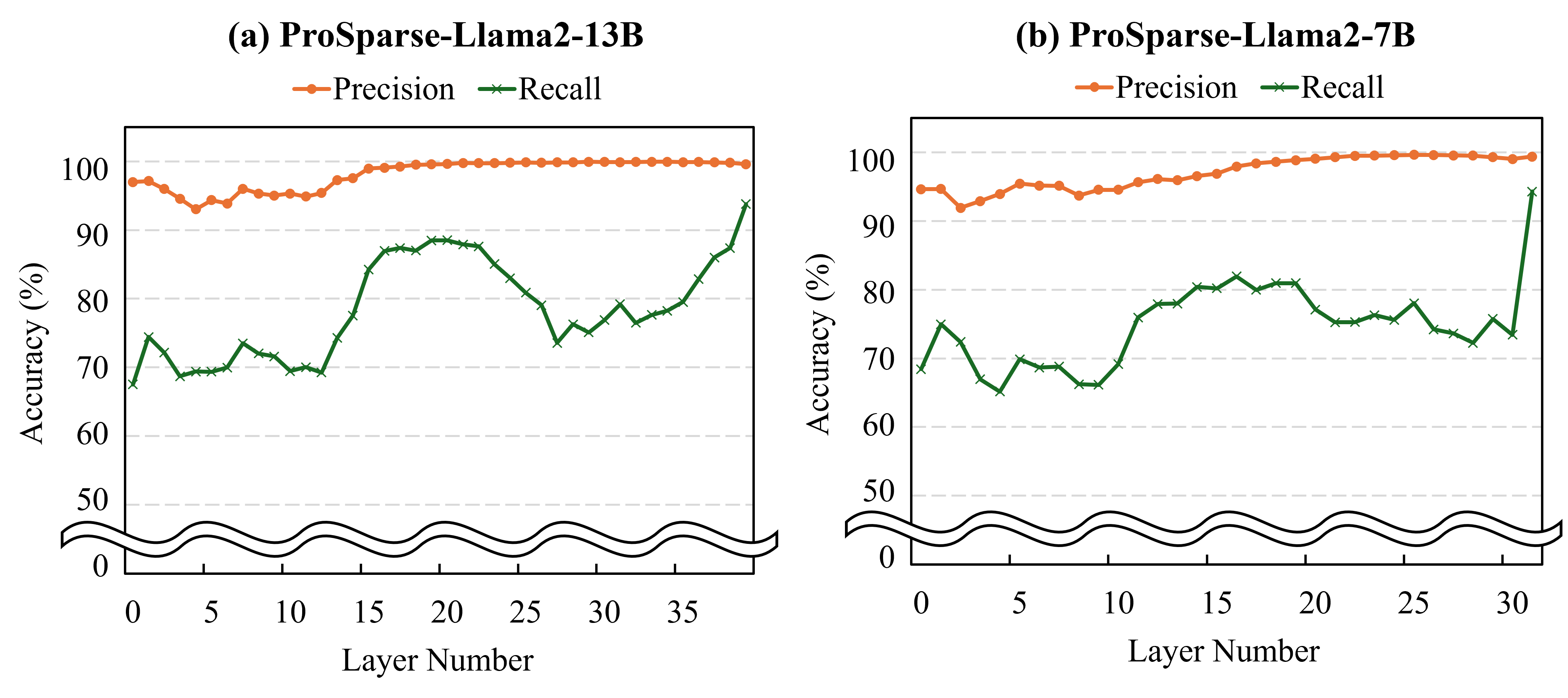}
\caption{Precision/Recall of \emph{SparseInfer}'s activation sparsity prediction method for each decoding layer of \emph{ProSparse-Llama2-7B} and \emph{ProSparse-Llama2-13B}.}
\label{fig:13Bscore}
\end{figure}

Fig.~\ref{fig:13Bscore} shows the precision and recall of the proposed activation sparsity prediction for each layer in the \emph{ProSparse-Llama2-7B} and \emph{ProSparse-Llama2-13B} models. 
Here, precision represents the proportion of elements in $h1$ that the predictor classified as sparse and are indeed sparse (i.e., the likelihood that an element predicted to be sparse is actually sparse). On the other hand, recall indicates the proportion of truly sparse elements in $h1$ that the predictor successfully identified as sparse, relative to the total number of sparse elements (i.e., how well the predictor captures the overall sparsity).
Overall, both models exhibited high precision values, reaching over 99\%
but a relatively lower precision was observed in the early layers compared to the later layers.

To compensate for the relatively lower accuracy in the early layers, the tunable parameter $\alpha$ can be applied differently across the layers. For the early layers where precision is relatively lower, where precision is relatively lower, $\alpha$ is set slightly above 1.0, while for the stabilized remaining layers, $\alpha$ is set to 1.0. The optimal value for $\alpha$ can be easily calibrated through test runs as the model changes, but empirically, setting $\alpha$ to a small value such as 1.01–1.03 for the early layers has shown good performance as will be shown in experiments. 

The proposed prediction method is training-free, making it robust to various standard quantization methods, e.g., INT8 or FP16. As long as the sign bit, i.e., MSB, can be extracted, it can be applied directly, regardless of the quantization scheme used. In contrast, training-based prediction schemes such as \emph{DEJAVU}~\cite{liu2023dejavu} would require re-training of the predictor to maintain its prediction accuracy whenever different quantization levels or methods are employed.

It should be emphasized that, as the proposed prediction scheme is training-free and tunable, it is adequate for the DSE of LLM inference which can find the optimal (time, accuracy) configurations for the given target platform and model.

\subsection{CUDA Implementation of SparseInfer}
\label{subsec:inference}

In this subsection, we present the implementation of the proposed method on a GPU using CUDA. 

\subsubsection{Pre-fetching and Packing Sign-Bit Information}
Extracting the sign bits by loading $W_{gate}$ at each layer during the decoding phase would result in unnecessary memory accesses. Instead, we pack the sign bits of 32 consecutive elements in $W_{gate}$ into a 32-bit integer variable when the model is loaded into memory. This process incurs a one-time overhead, as it is executed only once before \emph{llama.cpp} constructs the call graph. For the input vector $X$, which is determined dynamically, such pre-fetching is not feasible. Therefore, we pack the sign bits of its elements into 32-bit integer variables during the decoding phase, using a method similar to that used for $W_{gate}$.

\subsubsection{Sparsity Prediction}

CUDA employs a two-level hierarchical parallel structure, where a single kernel execution (grid) is composed of multiple thread blocks, and each thread block contains multiple threads. In our implementation for sparsity prediction in $X \cdot W_{gate}$,
each thread block is organized as a 2-dimensional $32 \times 16$ threads;
32 threads, or one warp, is assigned a row and 16 warps exist in a thread block. 
Consequently, the grid consists of $k/16$ thread blocks, 
where $k$ is the total number of rows in a matrix.
Note that, a warp in CUDA is an independent scheduling unit, and the synchronization, such as sum reduction, among the threads within a warp can be performed efficiently.
 
Listing 1 shows the CUDA kernel for activation sparsity prediction.
Each thread retrieves the 32-bit packed sign information from its corresponding position within $sign\_W\_gate$ and $sign\_X$, then performs an XOR operation to predict the sign of the product result (lines 6-8). Subsequently, the CUDA built-in function $\_\_popc()$ is utilized to count the number of ones in the XOR result, representing the number of elements predicted as negative, and this count is accumulated in $count$. The sign information of a single row of $W_{gate}$ is packed into a 32-bit format, requiring a total of $ncols = d/32$ XOR operations.
As explained above, a single warp is assigned a row and it needs to repeat lines 6-8 $ncols/WARP\_SIZE$ times to complete the counting (line 5).
Note that assigning more than one warp for the same row could reduce the number of iterations along the $ncols$, but it could require sum reduction among multiple warps.
Finally, it predicts an element to be sparse, setting the $skip$ flag for the corresponding row to true if the number of negative products ($count$) is larger than that of the positive products ($ncols*32 - count$). Note that, depending on a layer, a scale factor, $\alpha$, can be applied differently.

\begin{lstlisting}[caption={Acitvation Sparsity Prediction in CUDA}, label={code:predict}, numbers=left]
__global__ void predict_activation_sparsity(const int32_t *sign_W_gate, const int32_t *sign_X, int *skip, int nrows, int ncols, int alpha) {
    const int64_t row = (int64_t)blockIdx.x * blockDim.y + threadIdx.y;
    const int tid = threadIdx.x;
    int count = 0; 
    for (int i = 0; i < ncols / WARP_SIZE ; i++) {
        int col = i * WARP_SIZE + tid;
        int idx = row * ncols + col;
        int32_t Xor = sign_W[idx] ^ sign_X[col];
        count += __popc(Xor); } 
    count = warp_reduce_sum(count);  
    if (tid == 0) {
        if (count * 100 - (ncols * 32 - count) * alpha > 0) skip[row] = 0;
        else skip[row] = 1; }
}
\end{lstlisting}

\subsubsection{Sparse GEMV}

The sparse GEMV kernel takes as input the $skip$ flag for each row, which was predicted by Listing~\ref{code:predict}. 
The thread block size remains consistent with the Sparsity Predictor, set to $32 \times 16$. As before, each warp is assigned to handle one row and iteratively performs inner-product computations over $ncols$, accumulating the results. When the $skip$ flag is on, the warp immediately returns 0 without any computation. Since activation sparsity is determined at the row level and a sparse row results in the entire warp being skipped, there is no need for additional load balancing.
Note that, the sparse GEMV performed in the steps other than step 1 uses adjusted skip flags, which is the union of the predicted sparsity or previous flags \textit{and} the actual sparsity that can be obtained after step 1.

\subsubsection{Kernel Fusion}
Steps 1, 2, and 3 of the MLP operation described in Equation~(\ref{eq:mlp}) may optionally be fused into a single CUDA kernel. Since steps 1 and 2 share the input vector $X$, this kernel fusion enhances memory reusability. Additionally, step 3 uses the outputs of steps 1 and 2, making it efficient to include it in the same kernel. The activation sparsity prediction is performed before step 1, generating a $k$-length vector where sparse elements are marked with 1 and non-sparse elements with 0. Through this kernel fusion, memory access is limited to one load for $X$ and one write for the result of step 3, $h_3$. If each step were implemented as a separate kernel, as in the original llama.cpp implementation, $X$ would need to be loaded twice, while $h_1$ and $h_2$ would each require one load and one write, and $h_3$ would require one write. This illustrates the efficiency of the fused kernel approach.

On the other hand, step 4 is implemented as a separate kernel.
As $W_{down}$ is transposed to skip a row, instead of a column, the sum reduction of products in an inner-product in a GEMV cannot be implemented as accumulation but should be performed through $atomicAdd()$ among different warps, which are assigned different rows. If the flag indicates that the row is sparse, the warp skips the product computation and $atomicAdd()$, simply by returning. 
Note that, $W_{down}$ is transposed and stored as $W_{down}^T$ during model loading.

%% file: 6_experiments.tex
\section{Experimental Results}
\label{sec:exp}

The hardware platform used in the experiments is Jetson Orin 64GB~\cite{karumbunathan2022nvidia}, installed with Ubuntu 22.04, JetPack 6.0 DP, CUDA 12.2.
\emph{SparseInfer} was extended from \emph{llama.cpp}, implementing the activation sparsity prediction and exploitation explained in Section~\ref{sec:proposed}. Thus, the baseline for the comparison is \emph{llama.cpp}~\cite{llama_cpp}, and we also compared \emph{SparseInfer} with \emph{PowerInfer}~\cite{song2023powerinfer}, which is the state-of-the-art inference engine that supports activation sparsity. It is also based on \emph{llama.cpp} and employs the activation sparsity predictor of \emph{DEJAVU}~\cite{liu2023dejavu}.   
Assuming on-device environments, all the model weights were pre-loaded on GPU. \emph{PowerInfer} performed inference on GPU, not with a CPU-GPU hybrid execution. Note that Orin SoC integrates Cortex CPU and Ampere GPU in a single chip and they share the same LPDDR DRAM.
The models used are \emph{ProSparse-Llama2-13B} and \emph{7B}, which are ReLUfied Llama models. Note that \emph{SparseInfer} can perform inference with any ReLUfied models.

\begin{table}[b]
\scriptsize
\centering
\caption{Number of Operations for Prediction and MLP Block}
\begin{tabular}{c||c|c}
\hline
         & Prediction   & MLP Block   \\ \hline
\emph{llama.cpp} (dense) & 0                     & $2.123 \times 10^8$        \\ \hline
\emph{PowerInfer}    & $1.940 \times 10^7$      & $1.699 \times 10^7$        \\ \hline
\emph{SparseInfer} (proposed)      & $2.211 \times 10^6$      & $1.699 \times 10^7$        \\ \hline
\end{tabular}
\label{tab:computation_comparison}
\end{table}

\subsection{Prediction Overhead}
\label{subsec:dse_profile}

\subsubsection{Latency}

To compare the prediction overhead of \emph{SparseInfer} with \emph{PowerInfer}, the predictor latency with the first 10 samples in \emph{GSM8K} dataset was measured on these inference engines. \emph{PowerInfer} performs inference with \emph{PowerInfer-ProSparse-Llama2-13B} model where the weights of the \emph{DEJAVU} predictor trained with the rank size of 1024 is also included. Table~\ref{tab:computation_comparison} shows the number of operations in the \emph{PowerInfer} predictor is even larger than the actual computation amount for MLP block. \emph{SparseInfer} predictor has about an order of magnitude less number of operations than \emph{PowerInfer}, and note that the operation is 32-bit bitwise XOR in \emph{SparseInfer} while it is FP16 multiplication in \emph{PowerInfer}. As a result, the predictor latency of \emph{SparseInfe}r for one token generation per layer is only 70 usec on average, which is 3.66$\times$ faster than that of \emph{PowerInfer}. 
The speedup is lower than the operation reduction ratio because FP16 multiplications in \emph{PowerInfer} run on Tensor cores, while XORs run on CUDA cores.

\subsubsection{Memory Usage}

The hidden state dimension $d$ in \emph{ProSparse-Llama2-13B} is 5120 and $W_{gate}$ size in an MLP block is $d \times k$ = 5120$\times$13824, and the model has 40 MLP blocks. \emph{PowerInfer} requires (5120$\times$1024 + 1024$\times$13824)$\times$2(byte)$\times$40 = 1480 MB, when the rank in \emph{DEJAVU} predictor employed in \emph{PowerInfer} is 1024.
In contrast, \emph{SparseInfer} only requires the sign bit of each element, which is packed in 32-bit variables
,leading to 13824$\times$160$\times$4(byte)$\times$40 = 337.5 MB, which consumes 4.38$\times$ less memory usage.

\begin{figure}[t]
\centering
\includegraphics[width=.98\linewidth]
{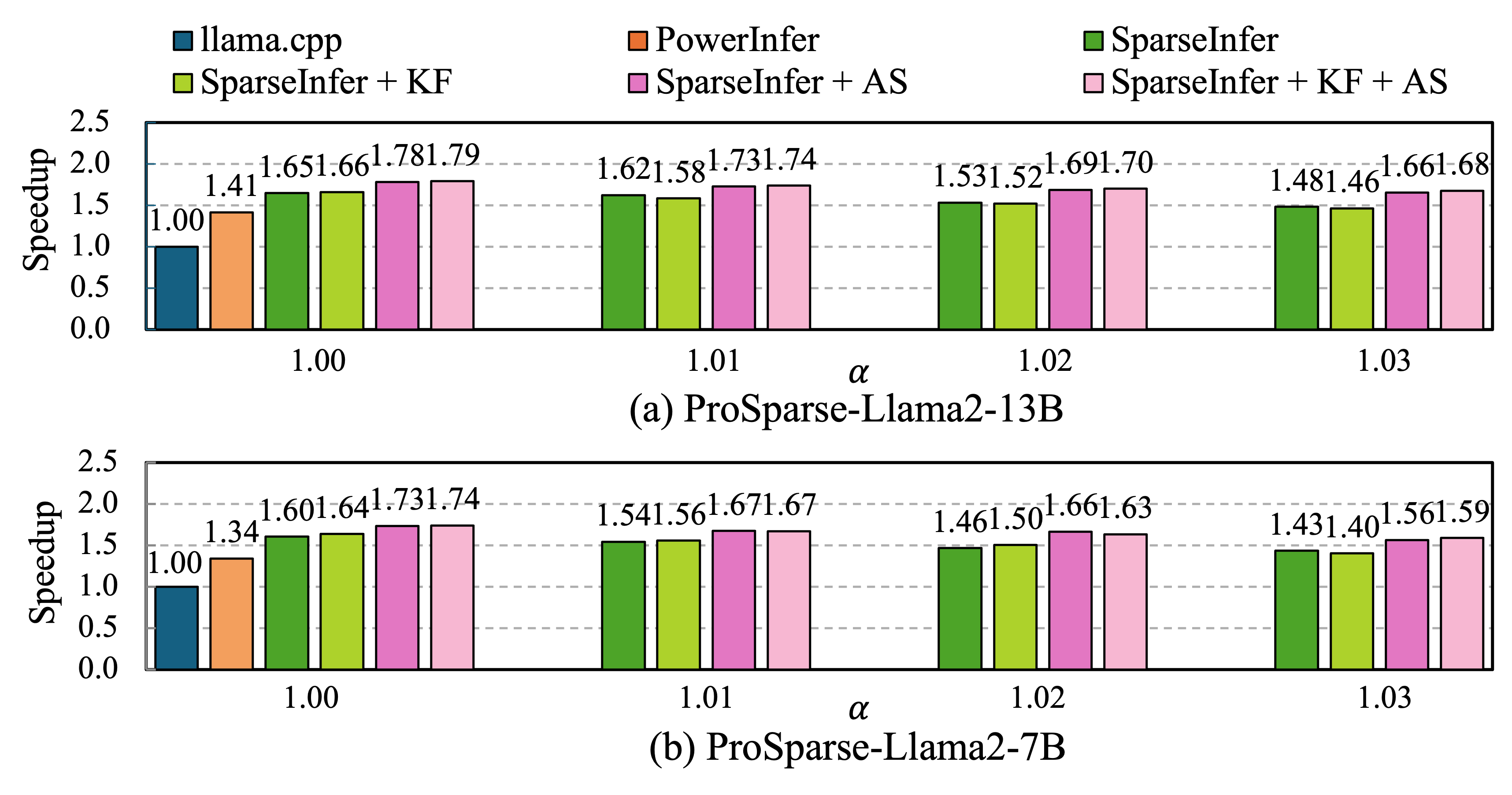}
\caption{Comparison of end-to-end latency for token generation.}
\label{fig:e2e_latency}
\end{figure}

\subsection{End-to-End Speedup}
\label{subsec:dse_space}

We measured the inference time of the first 10 samples in the \emph{GSM8K} dataset and compared the average end-to-end latency of token generation of \emph{SparseInfer} with \emph{PowerInfer}, as well as \emph{llama.cpp}, varying $\alpha$ from 1.00 to 1.03. Figure~\ref{fig:e2e_latency} shows the comparison results for \emph{ProSparse-Llama2-13B} and \emph{7B}, where fourfold \emph{SparseInfer} results were drawn depending on whether the kernel fusion was applied and whether the actual sparsity was exploited.

With $\alpha$ being 1.00, \emph{SparseInfer} achieved $1.79\times$ and $1.74\times$ speedups over \emph{llama.cpp} for 13B and 7B models, respectively, outperforming \emph{PowerInfer} by 1.27$\times$ and 1.30$\times$. As $\alpha$ gets larger, the speedups become slightly lower, due to the increased non-zero elements. 
With $\alpha$ being 1.03 for the 7B model, it leads to 1.59$\times$ over llama.cpp and 1.19$\times$ over \emph{PowerInfer}. 
We applied $\alpha>1.0$ only to the first 20 layers for both the 13B and 7B models.

The gain from the kernel fusion (\emph{+KF}) turned out to be insignificant 
due to the relatively small size of the input vector compared to the large cache size.
Utilizing \textit{actual sparsity} (\emph{+AS}), in addition to the predicted activation sparsity, contributes significantly to the speedup, especially when $\alpha$ gets larger as it conservatively predicts the sparsity.

\subsection{Accuracy}
\label{subsec:dse_ga}

To evaluate the effect of the proposed approach on model performance, lm-harness\cite{lm-eval}, a widely used benchmark suite was executed. As \emph{SparseInfer} utilizes activation sparsity only in the decoding phase, not in the prefill phase, log-likelihood-based benchmark evaluation is inadequate. We evaluated the model performance of \emph{ProSparse-Llama2-7B}, \emph{13B} with \emph{GSM8K}, \emph{BBH}~\cite{suzgunbbh2022} benchmarks that generate the answer.
Since a set of widely used benchmarks do not yet directly support \emph{llama.cpp}, we ported the \emph{SparseInfer} predictor and sparse GEMV to the PyTorch-based Transformer inference engine.
Additionally, we were unable to measure accuracy because \emph{PowerInfer} could not be ported to the Transformer engine.

The performance results are shown in Table~\ref{tab:performance_13B} and Table~\ref{tab:performance_7B}. Without scaling for the earlier layers, the performance drop occurs, especially in \emph{ProSparse-Llama2-7B}. However, it becomes negligible within 1\%p when $\alpha$ = 1.03 is applied. 
Note that random selection with the 90\% activation sparsity, instead of the prediction, resulted in 0\% accuracy.

\begin{table}[t]
\scriptsize
\centering
\caption{Performance comparison for \emph{ProSparse-Llama2-13B}}
\begin{tabular}{c|c||c|c|c}
\hline
\textbf{Method}         & \textbf{$\alpha$}   & \textbf{GSM8K}   & \textbf{BBH}   & \textbf{Average}   \\ \hline\hline
\textbf{Baseline}       & -                       & 30.71            & 44.80          & 37.76              \\ \hline
\multirow{4}{*}{\textbf{\emph{SparseInfer}}} 
                        & 1.00                    & 27.75 (-2.96)           & 42.91 (-1.89)         & 35.33 (-2.43)              \\ \cline{2-5} 
                        & 1.01                    & 28.73  (-1.98)          & 43.57 (-1.23)              & 36.15 (-1.61)                  \\ \cline{2-5} 
                        & 1.02                    & 29.64 (-1.07)           & 44.43 (-0.37)         & 37.04 (-0.72)              \\ \cline{2-5} 
                        & 1.03                    & 30.63 (-0.08)            & 44.34 (-0.46)         & 37.49 (-0.27)              \\ \hline
\end{tabular}
\label{tab:performance_13B}
\end{table}

\begin{table}[t]
\scriptsize
\centering
\caption{Performance comparison for \emph{ProSparse-Llama2-7B}}
\begin{tabular}{c|c||c|c|c}
\hline
\textbf{Method}         & \textbf{$\alpha$}   & \textbf{GSM8K}   & \textbf{BBH}   & \textbf{Average}   \\ \hline\hline
\textbf{Baseline}       & -                       & 13.42            & 35.80          & 24.61              \\ \hline
\multirow{4}{*}{\textbf{\emph{SparseInfer}}} 
                        & 1.00                    &  8.95 (-4.47)             & 27.37 (-8.43)         & 18.16 (-6.45)             \\ \cline{2-5} 
                        & 1.01                    & 10.77 (-2.65)           & 33.70 (-2.10)         & 22.24 (-2.37)             \\ \cline{2-5} 
                        & 1.02                    & 11.30 (-2.12)           & 35.51 (-0.29)         & 23.41 (-1.20)             \\ \cline{2-5} 
                        & 1.03                    & 12.96 (-0.46)           & 35.60 (-0.20)         & 24.28 (-0.33)             \\ \hline
\end{tabular}
\label{tab:performance_7B}
\end{table}

%% file: 7_conclusion.tex
\section{Conclusion}
\label{sec:conclusion}
This paper proposes an efficient yet sufficiently accurate prediction method for activation sparsity in LLMs, by which the memory and computation in the inference can be significantly reduced. Based on the observation of the input activation and weight distribution, the prediction method requires only sign-bit information of the weight matrix and the input vector, making the approach training-free and DSE-friendly.
\textit{SparseInfer}, a framework for fast LLM inference, employing the proposed predictor, allows designers to explore the different trade-offs of LLM inference, given the target architecture and the model. The experimental results show that \textit{SparseInfer} outperforms \textit{PowerInfer}, the state-of-the-art, by 21\% while maintaining the minimal accuracy loss on Jetson Orin AGX.

%% file: 00_main.bbl
\begin{thebibliography}{10}
\providecommand{\url}[1]{#1}
\csname url@samestyle\endcsname
\providecommand{\newblock}{\relax}
\providecommand{\bibinfo}[2]{#2}
\providecommand{\BIBentrySTDinterwordspacing}{\spaceskip=0pt\relax}
\providecommand{\BIBentryALTinterwordstretchfactor}{4}
\providecommand{\BIBentryALTinterwordspacing}{\spaceskip=\fontdimen2\font plus
\BIBentryALTinterwordstretchfactor\fontdimen3\font minus \fontdimen4\font\relax}
\providecommand{\BIBforeignlanguage}[2]{{%
\expandafter\ifx\csname l@#1\endcsname\relax
\typeout{** WARNING: IEEEtran.bst: No hyphenation pattern has been}%
\typeout{** loaded for the language `#1'. Using the pattern for}%
\typeout{** the default language instead.}%
\else
\language=\csname l@#1\endcsname
\fi
#2}}
\providecommand{\BIBdecl}{\relax}
\BIBdecl

\bibitem{radford2018improving}
A.~Radford, ``Improving language understanding by generative pre-training,'' 2018.

\bibitem{vaswani2017attention}
A.~Vaswani, ``Attention is all you need,'' \emph{Advances in Neural Information Processing Systems}, 2017.

\bibitem{ma2023llm}
X.~Ma, G.~Fang, and X.~Wang, ``Llm-pruner: On the structural pruning of large language models,'' \emph{Advances in neural information processing systems}, vol.~36, pp. 21\,702--21\,720, 2023.

\bibitem{yao2022zeroquant}
Z.~Yao, R.~Yazdani~Aminabadi, M.~Zhang, X.~Wu, C.~Li, and Y.~He, ``Zeroquant: Efficient and affordable post-training quantization for large-scale transformers,'' \emph{Advances in Neural Information Processing Systems}, vol.~35, pp. 27\,168--27\,183, 2022.

\bibitem{ramachandran2017searching}
P.~Ramachandran, B.~Zoph, and Q.~V. Le, ``Searching for activation functions,'' \emph{arXiv preprint arXiv:1710.05941}, 2017.

\bibitem{hendrycks2016gaussian}
D.~Hendrycks and K.~Gimpel, ``Gaussian error linear units (gelus),'' \emph{arXiv preprint arXiv:1606.08415}, 2016.

\bibitem{mirzadeh2023relufication}
I.~Mirzadeh, K.~Alizadeh, S.~Mehta, C.~C.~D. Mundo, O.~Tuzel, G.~Samei, M.~Rastegari, and M.~Farajtabar, ``Relu strikes back: Exploiting activation sparsity in large language models,'' \emph{arXiv preprint arXiv:2310.04564}, 2023.

\bibitem{song2024prosparse}
C.~Song, X.~Han, Z.~Zhang, S.~Hu, X.~Shi, K.~Li, C.~Chen, Z.~Liu, G.~Li, T.~Yang, and M.~Sun, ``Prosparse: Introducing and enhancing intrinsic activation sparsity within large language models,'' \emph{arXiv preprint arXiv:2402.13516}, 2024.

\bibitem{liu2023dejavu}
Z.~Liu, J.~Wang, T.~Dao, T.~Zhou, B.~Yuan, Z.~Song, A.~Shrivastava, C.~Zhang, Y.~Tian, C.~Re \emph{et~al.}, ``Deja vu: Contextual sparsity for efficient llms at inference time,'' in \emph{International Conference on Machine Learning}.\hskip 1em plus 0.5em minus 0.4em\relax PMLR, 2023, pp. 22\,137--22\,176.

\bibitem{llama_cpp}
G.~Gerganov, ``llama.cpp,'' \url{https://github.com/ggerganov/llama.cpp}, 2023, accessed: September 18th, 2024.

\bibitem{kurtz2020inducing}
M.~Kurtz, J.~Kopinsky, R.~Gelashvili, A.~Matveev, J.~Carr, M.~Goin, W.~Leiserson, S.~Moore, N.~Shavit, and D.~Alistarh, ``Inducing and exploiting activation sparsity for fast inference on deep neural networks,'' in \emph{International Conference on Machine Learning}.\hskip 1em plus 0.5em minus 0.4em\relax PMLR, 2020, pp. 5533--5543.

\bibitem{relullama}
{ReluLLaMa}, ``{ReluLLaMa},'' \url{https://huggingface.co/SparseLLM/ReluLLaMA-7B}.

\bibitem{song2023powerinfer}
Y.~Song, Z.~Mi, H.~Xie, and H.~Chen, ``Powerinfer: Fast large language model serving with a consumer-grade gpu,'' \emph{arXiv preprint arXiv:2312.12456}, 2023.

\bibitem{song2024turbo}
Y.~Song, H.~Xie, Z.~Zhang, B.~Wen, L.~Ma, Z.~Mi, and H.~Chen, ``Turbo sparse: Achieving llm sota performance with minimal activated parameters,'' \emph{arXiv preprint arXiv:2406.05955}, 2024.

\bibitem{liu2024teal}
J.~Liu, P.~Ponnusamy, T.~Cai, H.~Guo, Y.~Kim, and B.~Athiwaratkun, ``Training-free activation sparsity in large language models,'' \emph{arXiv preprint arXiv:2408.14690}, 2024.

\bibitem{cao2019seernet}
S.~Cao, L.~Ma, W.~Xiao, C.~Zhang, Y.~Liu, L.~Zhang, L.~Nie, and Z.~Yang, ``Seernet: Predicting convolutional neural network feature-map sparsity through low-bit quantization,'' in \emph{Proceedings of the IEEE/CVF Conference on Computer Vision and Pattern Recognition}, 2019, pp. 11\,216--11\,225.

\bibitem{touvron2023llama}
H.~Touvron, T.~Lavril, G.~Izacard, X.~Martinet, M.-A. Lachaux, T.~Lacroix, B.~Rozi{\`e}re, N.~Goyal, E.~Hambro, F.~Azhar \emph{et~al.}, ``Llama: Open and efficient foundation language models,'' \emph{arXiv preprint arXiv:2302.13971}, 2023.

\bibitem{almazrouei2023falcon}
E.~Almazrouei, H.~Alobeidli, A.~Alshamsi, A.~Cappelli, R.~Cojocaru, M.~Alhammadi, M.~Daniele, D.~Heslow, J.~Launay, Q.~Malartic \emph{et~al.}, ``The falcon series of language models: Towards open frontier models,'' \emph{Hugging Face repository}, 2023.

\bibitem{zhang2023opt}
S.~Zhang, S.~Roller, N.~Goyal, M.~Artetxe, M.~Chen, S.~Chen, C.~Dewan, M.~Diab, X.~Li, X.~V. Lin \emph{et~al.}, ``Opt: Open pre-trained transformer language models, 2022,'' \emph{URL https://arxiv. org/abs/2205.01068}, vol.~3, pp. 19--0, 2023.

\bibitem{tim2023qlora}
T.~Dettmers, A.~Pagnoni, A.~Holtzman, and L.~Zettlemoyer, ``Qlora: efficient finetuning of quantized llms,'' in \emph{Proceedings of the 37th International Conference on Neural Information Processing Systems}, ser. NIPS '23.\hskip 1em plus 0.5em minus 0.4em\relax Red Hook, NY, USA: Curran Associates Inc., 2024.

\bibitem{Craig1936OnTF}
\BIBentryALTinterwordspacing
C.~C. Craig, ``On the frequency function of \$xy\$,'' \emph{Annals of Mathematical Statistics}, vol.~7, pp. 1--15, 1936. [Online]. Available: \url{https://api.semanticscholar.org/CorpusID:121925566}
\BIBentrySTDinterwordspacing

\bibitem{Karlgms8k}
C.~Karl, V.~Kosaraju, M.~Bavarian, M.~Chen, H.~Jun, L.~Kaiser, M.~Plappert \emph{et~al.}, ``Training verifiers to solve math word problems,'' \emph{arXiv preprint arXiv:2110.14168}, 2021.

\bibitem{karumbunathan2022nvidia}
L.~S. Karumbunathan, ``Nvidia jetson agx orin series,'' 2022.

\bibitem{lm-eval}
EleutherAI, ``lm-evaluation-harness,'' \url{https://github.com/EleutherAI/lm-evaluation-harness}, 2024, accessed: September 23th, 2024.

\bibitem{suzgunbbh2022}
M.~Suzgun, N.~Scales, N.~Schärli, S.~Gehrmann, Y.~Tay, H.~W. Chung, A.~Chowdhery, Q.~V. Le, E.~H. Chi, D.~Zhou \emph{et~al.}, ``Challenging big-bench tasks and whether chain-of-thought can solve them,'' \emph{arXiv preprint arXiv:2210.09261}, 2022.

\end{thebibliography}
